\documentclass[10pt,a4paper]{article}
\pdfoutput=1
\usepackage{jheppub}

\usepackage{amssymb}
\usepackage{amsmath}
\usepackage[utf8]{inputenc}
\usepackage{graphicx}
\usepackage{bbm}

\def\Eq#1{Eq.~(\ref{#1})}

\def\Sec#1{Sec.~\ref{#1}}

\def\p{{\bf p}}

\begin{document}

\title{Global fluid fits to identified particle transverse momentum spectra from heavy-ion collisions at the Large Hadron Collider}

\author[a]{D. Devetak,}
\emailAdd{d.devetak@cern.ch}
\affiliation[a]{Physikalisches Institut, Universit{\"a}t Heidelberg, 69120 Heidelberg, Germany}

\author[b,c]{A. Dubla,}
\emailAdd{a.dubla@cern.ch}
\affiliation[b]{GSI Helmholtzzentrum f{\"u}r Schwerionenforschung, 64291 Darmstadt, Germany}
\affiliation[c]{Nikhef National Institute for Subatomic Physics, Amsterdam, The Netherlands}

\author[d]{S. Floerchinger,}
\emailAdd{stefan.floerchinger@thphys.uni-heidelberg.de}
\affiliation[d]{Institut f\"{u}r Theoretische Physik, Universit\"{a}t Heidelberg, 69120 Heidelberg, Germany} 
\author[d]{E. Grossi}
\emailAdd{e.grossi@thphys.uni-heidelberg.de}

\author[a,b]{S.~Masciocchi,}
\emailAdd{s.masciocchi@gsi.de}

\author[e]{A. Mazeliauskas}
\emailAdd{aleksas.mazeliauskas@cern.ch}
\affiliation[e]{Theoretical Physics Department, CERN, CH-1211 Gen\`eve 23, Switzerland} 

\author[b]{and I. Selyuzhenkov}
\emailAdd{ilya.selyuzhenkov@gmail.com}

\date{\today}

\abstract{
Transverse momentum spectra of identified particles produced in heavy-ion collisions at the Large Hadron Collider are described with relativistic fluid dynamics. We perform a systematic comparison of experimental data for pions, kaons and protons up to a transverse momentum of 3 GeV$/c$ with calculations using the \textsc{Fluid{\it u}M} code package to solve the evolution equations of fluid dynamics, the \textsc{TrENTo} model to describe the initial state and the \textsc{FastReso} code to take resonance decays into account. Using data in five centrality classes at the center-of-mass collision energy per nucleon pair $\sqrt{s_\text{NN}}=2.76\,\text{TeV}$, we determine systematically the most likely parameters of our theoretical model including the shear and bulk viscosity to entropy ratios, the initialization time, initial density and freeze-out temperature through a global search and quantify their posterior probability. This is facilitated by the very efficient numerical implementation of \textsc{Fluid{\it u}M} and \textsc{FastReso}. Based on the most likely model parameters we present predictions for the transverse momentum spectra of multi-strange hadrons as well as identified particle spectra from Pb--Pb collisions at $\sqrt{s_\text{NN}}=5.02\,\text{TeV}$.
}

\maketitle

\section{Introduction}

High-energy heavy-ion collisions at the Relativistic Heavy Ion Collider (RHIC)
and the Large Hadron Collider (LHC) produce a fluid consisting of quarks and
gluons, the fundamental constituents of Quantum Chromodynamics (QCD)~\cite{Busza:2018rrf,Aamodt:2010pa, ADAMS2005102, ADCOX2005184}.
The produced fluid is particularly
interesting because it is described on microscopic level by a renormalizable
and fundamental quantum field theory.
While first principle calculations of the macroscopic fluid properties are challenging,
phenomenological and theoretical studies are motivated by an increasing amount of experimental results.
Remarkably, data and  models suggest that a fluid dynamic expansion 
might be behind some of the recent results of collective behaviour in yet smaller proton-nucleus and proton-proton systems~\cite{Nagle:2018nvi,PHENIX:2018lia, Acharya:2019vdf}. 
Alternative descriptions in terms of initial-state physics and medium-less hadron production are also being developed~\cite{Nagle:2018nvi, Mace:2018vwq, Bierlich:2018xfw, Greif:2017bnr, Lin:2004en, Zhou:2015iba},
all of which questions the uniqueness of a fluid-like response of the quark-gluon plasma (QGP).
Therefore the resolution to the origins of collective behaviour will likely rely on quantitative rather than just qualitative agreement between data and model. 
To this end, we present a new framework for systematic studies of
soft hadronic observables based on up-to-date and efficient
modelling of heavy-ion collisions.

We combine the successful initial condition model \textsc{TrENTo}~\cite{Moreland:2014oya},
with the recent viscous relativistic fluid dynamics implementation
\textsc{Fluid{\it u}M}~\cite{Floerchinger:2018pje} 
and the novel resonance decay procedure \textsc{FastReso}~\cite{Mazeliauskas:2018irt}. The mode
splitting implemented in \textsc{Fluid{\it u}M} allows for a very fast evolution
with a single event taking mere seconds to compute. In our
work we use an equation of state $p(T)$ based on recent
Lattice-QCD calculations~\cite{Borsanyi:2016ksw,Bazavov:2014pvz}, see
\cite{Floerchinger:2018pje}, and include both shear and bulk viscous corrections
in the evolution and particle freeze-out. In addition we use an enlarged set of resonance decays
~\cite{Alba:2017mqu, Alba:2017hhe, parotto_private} based on the 2016 edition of the
 Particle Data Group book~\cite{Patrignani:2016xqp}. 

In the absence of precise first principle calculations, the
phenomenological description of heavy-ion collisions has a
number of open parameters at different stages of the  evolution.
They can be estimated indirectly from the comparison of simulations to experimental results. 
Obviously, a too
large number of such parameters can limit the precision of this estimate.
Moreover,
covering a multi-dimensional parameter space is computationally expensive and requires
efficient implementation of the model.
In the present work, we determine the specific shear and bulk viscosities of the QGP,
as well as the freeze-out temperature $T_\text{fo}$, the starting time of a fluid description $\tau_0$ 
and the initial entropy profile normalization.

Previous multi-observable model-to-data fits focused on the 
centrality dependence of momentum integrated quantities, like particle multiplicity, mean
transverse momentum or flow harmonics~\cite{Bernhard:2016tnd}.
In this work we 
perform a systematic study of more differential data, namely, transverse momentum spectra with $p_\mathrm{T}<$ 3 GeV/$c$ of
$\pi$, $K$, and $p$ in five centrality classes of Pb--Pb collisions at the center-of-mass energy per nucleon pair $\sqrt{s_\text{NN}}=2.76\,\text{TeV}$ at the LHC,
and compare them with fluid dynamic simulations.

Let us mention here already that we are able to put interesting and non-trivial constraints on transport properties, specifically shear and bulk viscosity, from the analysis of transverse momentum spectra for identified particles alone. This might come as a surprise to some readers because it was believed so far that significant constraints on transport properties need an analysis of anisotropic flow. While flow coefficients are indeed expected to contain even more detailed information, we want to emphasize that experimental data on transverse momentum spectra are by now of a rather high quality. We can exploit this here and perform a detailed statistical analysis including fits with systematic $\chi^2$ minimization. We find that $\chi^2$ rises rather quickly away from the global minimum which leads to surprisingly tight constraints on the QCD fluid properties.

We summarize the details of initial condition, evolution and hadronization procedures in 
\Sec{sec:setup}. We discuss the fit procedure and determination of its
uncertainties in \Sec{sec:analysis}. We then provide the best fit results and predictions
for additional observables in \Sec{sec:spectra} and \Sec{sec:predictions}. Finally, we discuss the analysis and future directions in \Sec{sec:discussion}.

\section{Setup\label{sec:setup}}
In this section we briefly describe the different components of our theoretical model. We start with the time evolution as implemented in \textsc{Fluid{\it u}M} \cite{Floerchinger:2018pje} solving the equations of relativistic fluid dynamics with shear and bulk viscosity and corresponding relaxation times. Subsequently we turn to the initial conditions, specifically the shape of the energy density in the transverse plane for which we use the \textsc{TrENTo} model \cite{Moreland:2014oya}. 
Finally, kinetic freeze-out and the implementation of strong resonance decays is done using \textsc{FastReso} \cite{FastReso}.

\subsection{Hydrodynamic evolution: \textsc{Fluid{\it u}M}}

To solve the relativistic fluid equations of motion, we use the code package \textsc{Fluid{\it u}M} \cite{Floerchinger:2018pje}. It is based on the theoretical framework of relativistic fluid dynamics with mode expansion \cite{Floerchinger:2013rya, Floerchinger:2013hza, Floerchinger:2014fta}, where the fluid dynamic fields are decomposed in terms of a background-fluctuation splitting, similar to what is done for example in cosmology. Schematically, we write the fluid fields $\Phi(\tau ,r,\phi ,\eta )={ \Phi  }_{ 0 }(\tau ,r)+{ \Phi }_{ 1 }(\tau ,r,\phi ,\eta)$. The non-linear evolution equations for an azimuthally and Bjorken boost symmetric background ${ \Phi  }_{ 0 }(\tau ,r)$ are solved first, while azimuthally and rapidity dependent perturbations ${ \Phi }_{ 1 }(\tau ,r,\phi ,\eta)$ around this are then studied separately. The evolution equations for both the background and the perturbations around them can be implemented with very accurate and highly efficient numerical algorithms \cite{Floerchinger:2018pje}. 

For the present paper we are interested in azimuthally averaged transverse momentum spectra of identified particles in the mid-rapidity region and do not consider azimuthally and rapidity-dependent perturbations. Neglecting terms that are of quadratic or higher order in perturbation amplitudes, we need only the background solution to the fluid evolution equations as calculated from \textsc{Fluid{\it u}M}. The corresponding equations of motion have been analyzed from a mathematical perspective, and with an emphasis on their causality structure in ref.\ \cite{Floerchinger:2017cii}. 

We note here that the current implementation of \textsc{Fluid{\it u}M} features a flow at vanishing net baryon number chemical potential, based on a state-of-the-art thermodynamic equation of state~\cite{Borsanyi:2016ksw,Bazavov:2014pvz}, as well as shear and bulk viscous dissipation. For the present paper we assume the shear viscosity to entropy ratio $\eta/s$ to be independent of temperature. The bulk viscosity to entropy ratio $\zeta / s$ is taken to be temperature dependent, however. Specifically, we assume it to be of the Lorentzian form
\begin{equation}
\label{eq:bulk}
  \zeta/s= \frac{\left(\zeta/ s\right)_{\text{max}}}{1+\left(\frac{T-T_{\text{peak}}}{\Delta T}\right)^2},
\end{equation}
with the peak temperature $T_{\text{peak}}=175$ MeV and $\Delta T=24$ MeV \cite{Moreland:2018gsh}. The maximum value $(\zeta/ s)_{\text{max}}$ is taken as a fit parameter.

Shear and bulk relaxation times are assumed to be determined by the relations~\cite{Denicol:2014vaa}
\begin{equation}
\label{eq:Taupi}
  \frac{\tau_\text{shear}}{\eta/(\epsilon+p)}=
  5 ,\qquad
  \frac{\tau_\text{bulk}}{\zeta/(\epsilon+p)} = \frac{1}{15\left( \frac{1}{3}- c_s^2 \right)^2 }+\frac{a}{\zeta/(\epsilon+p)},
\end{equation}
where $\epsilon$ is the energy density, $p$ is the pressure, $c_s$ is the (temperature dependent) velocity of sound, and $a=0.1\, \text{fm/c}$ is a small offset such that a causal evolution of the radial expansion is indeed ensured \cite{Floerchinger:2017cii}. For more details on the implementation we refer to \cite{Floerchinger:2018pje}. 

\subsection{Initial conditions: \textsc{TrENTo}}

In general terms, a characterization of the initial conditions for Israel-Stewart type fluid dynamics with azimuthal rotation and longitudinal boost symmetry as used for the background in \textsc{Fluid{\it u}M} consists of the temperature $T$, radial fluid velocity $u^r$, two independent components of shear stress $\pi_{\phi}^{\phi}$ and $\pi_{\eta}^{\eta}$ as well as bulk viscous pressure $\pi_{\mathrm{bulk}}$ on some initial Cauchy surface, such as $\tau=\tau_0$. In the present work we neglect initial radial flow and assume initially $\pi_{\phi}^{\phi}=\pi_{\eta}^{\eta}=\pi_{\mathrm{bulk}}=0$. This choice is respecting relativistic causality \cite{Floerchinger:2017cii}.

The shape of the initial entropy density distribution in the transverse plane (which determines the temperature through the thermodynamic equation of state) is taken from the initial state model \textsc{TrENTo} \cite{Moreland:2014oya}, with an overall normalization factor that we take as a fit parameter.
The parameters of \textsc{TrENTo} have been taken as in ref.\ \cite{Moreland:2014oya}, in particular we selected the reduced thickness parameter $p=0$,  the fluctuation parameter $k=1.4$, the nucleon width $\sigma=0.6 \text{ fm}$ and the inelastic nucleon-nucleon cross section $\sigma^{\text{NN}}_{\text{inel}}=6.4 \text{ fm}^2$. Using this set of parameters we have generated the transverse density $T_\text{R}(x,y)$ for $10^5$ events with impact parameter sampled from the range $b\in[0 \, \text{fm} , 20 \, \text{fm}]$ and randomized event plane angle. As usual, the distribution of impact parameters is governed by the random distribution of nuclei in the transverse plane and the probability for them to scatter in the \textsc{TrENTo} model.  It is convenient to shift the events in the transverse plane such that $\int d^2 x \{ \vec x \, T_\text{R}(\vec x) \}= 0$.

The integrated transverse density $\int d^2x  T_\text{R}(\vec x)$ is expected to be monotonously related to the total final charged particle multiplicity, therefore we used this quantity to divide the generated events into narrow multiplicity classes of one percent. Each of these centrality classes can be seen as an ensemble of events with random orientation in the transverse plane.

For each centrality class we calculate the averaged or expected entropy density profile as 
\begin{equation}
\label{eq:sinitial}
    s(r)=\frac{\text{Norm}_i}{\tau_0}
    \left\langle T_\text{R}(r,\phi) \right\rangle.
\end{equation}
(Note that for ensembles with random orientation in the transverse plane the right hand side is independent of $\phi$.) We introduce here a normalization constant  $\text{Norm}_i$ for each centrality class $i$.
Ideally, the initial state model should take care of centrality dependence and all centrality classes would have one identical normalization. While the parameter choice $p=0$ in \textsc{TrENTo} comes close to this, we observe some residual tension with the data which we lift by allowing the normalization to be centrality class dependent.
We have taken out the  initialization time $\tau_0$ to already take into account the main effect of the longitudinal expansion (Bjorken flow) at early times. The initial temperature as a function of radius is then obtained using the equation of state. 

While it is convenient for the theoretical description to work with rather narrow centrality classes, they are typically somewhat larger in the experimental results. There are now two possible strategies to deal with this. The first would be to calculate particle spectra for each of the narrow classes and to combine (average) them in a convenient way afterwards. The second strategy is to produce averaged entropy densities for the larger centrality classes by averaging the corresponding distributions from the more narrow classes and to propagate those. The difference in experimental observables between both procedures can be taken as an estimate for the importance of fluctuations. We have compared both strategies and found the difference for transverse momentum spectra to be rather small, of the order of $1\%$ for central collisions. Because of the advantage with respect to computational costs, we follow therefore the second strategy.

\subsection{Freeze-out and resonance decays: \textsc{FastReso}}

As the system cools down and dilutes, it crosses from a quark-gluon plasma to a fluid dominated by hadronic degrees of freedom. The fluid dynamic description of the latter breaks down eventually, because particle scatterings are no longer efficient
in maintaining (first chemical and then kinetic) equilibrium. This necessitates
the conversion of fluid fields, such as temperature and flow velocity, to
the distribution of hadronic degrees of freedom. 

The dynamics of hadronization is not completely understood, but lattice QCD
calculations show that below the QCD pseudo-critical temperature
$T_{pc}=156\pm1.5\,\text{MeV}$~\cite{Bazavov:2018mes,Steinbrecher:2018phh},
color neutral hadrons become the dominant degrees of freedom of the plasma.
In particular the equation of state approaches that of a hadron resonance gas
(HRG)~\cite{Alba:2017hhe}. 

Around or somewhat below $155\,\text{MeV}$ in temperature, fluid fields are customary converted to particle distributions using
Cooper-Frye procedure
~\cite{Cooper:1974mv}. The spectrum of hadron species $a$ on the freeze-out hypersurface $\Sigma$ is given by the following integral
\begin{equation}
E_\p\frac{d N_a}{d^3 \p} = \frac{\nu_a}{(2\pi)^3}\int_\Sigma f_a(\bar{E}_\p) 
p^\mu d\Sigma_\mu,\label{eq:CF}
\end{equation}
where $\nu_a$ is the degeneracy factor of spin/polarization states and $f_a$ is
a particle distribution function, which, in addition to the particle energy in fluid
rest-frame $\bar{E}_\p \equiv-u^\nu p_\nu$, may also depend on the local
temperature $T(x)$, fluid velocity $u^\mu(x)$, chemical potential $\mu(x)$,
viscous shear-stress $\pi^{\mu\nu}(x)$ and bulk viscous pressure $\pi_\text{bulk}(x)$.

Chemical freeze-out takes place when particle species changing interactions
are no longer able to keep up with the expansion rate. However, in practice, a simpler criterion based
solely on the freeze-out temperature is used and the freeze-out surface
$\Sigma$ is assumed to be a surface of constant temperature. One sometimes includes after chemical freeze-out and before kinetic freeze-out a phase described by fluid dynamics but for a fluid in partial chemical equilibrium (see \cite{Bebie:1991ij} for pioneering work in this regard). Such a fluid is governed by a number of conservation laws in addition to the ones for energy and momentum. We have implemented this in our theoretical model but found eventually that the improvement of transverse momentum spectra of the studied particles species 
is not significant. For this reason we use in the present work a simpler prescription with only a single, chemical {\it and } kinetic freeze-out.

On the freeze-out surface we take the particle distribution function to be given by the equilibrium
Bose-Einstein or Fermi-Dirac distribution (depending on the species), modified by additional corrections due to bulk and shear viscous dissipation, 
\begin{equation}
  f = f_\text{eq} + \delta f^\text{bulk} + \delta f^\text{shear}. \label{eq:initialf}
\end{equation}
For the viscous corrections we use the commonly employed parametrizations~\cite{Teaney:2003kp,Paquet:2015lta} 
\begin{align}
&\delta f^\text{bulk} = f_\text{eq}(1\pm f_\text{eq})\left[\frac{\bar E_p}{T}\left(\tfrac{1}{3}-c_s^2\right)-\frac{m^2}{3 T\bar E_p} \right]\frac{\pi_\text{bulk}}{\zeta/\tau_\text{bulk}},\label{eq:bulkini}\\
&\delta f^\text{shear} = f_\text{eq}(1\pm f_\text{eq})\frac{\pi_{\rho\nu}p^\rho p^\nu}{2(\epsilon+p)T^2}.\label{eq:shearini}
\end{align}
Here $m$ is the mass of the primary resonance.

After freeze-out the populations of unstable resonances decrease as a consequence of their decays and feed the spectra of long lived particles. This large modification of the
pion, kaon and proton spectra can be calculated by decaying all (sufficiently unstable) resonances. An efficient procedure to calculate these direct decays was recently introduced by some of us in ref.~\cite{Mazeliauskas:2018irt}. The main idea is to apply the decay maps to the primary distributions in \Eq{eq:CF} \emph{before} doing
the surface integral. The resulting distribution function of decay products can be decomposed into irreducible components (with respect to rotations in the fluid rest frame) that are pre-computed and stored~\cite{FastReso}. Furthermore, for the case of azimuthally symmetric and boost-invariant surface, the freeze-out integrals over space-time rapidity and azimuthal angle can also be pre-computed. Parametrizing the remaining 1+1 dimensional freeze-out surface in radial coordinates
by $(\tau(\alpha), r(\alpha) )$ where $\alpha\in (0,1)$ is some parameter, the Cooper-Frye freeze-out 
integral simplifies to one-dimensional integral over $\alpha$,
\begin{align}
\frac{d N}{2\pi p_Tdp_Tdy} = & \frac{\nu}{(2\pi)^3} \int_0^1\! d\alpha \; \tau(\alpha) r(\alpha)  
\nonumber\\
& \times {\Bigg \{} \frac{\partial r}{\partial\alpha}
\Big[
K^\text{eq}_1 + \frac{\pi^\eta_\eta}{2(\epsilon+p)T^2} \; K_1^\text{shear}
+ \frac{\pi^\phi_\phi}{2(\epsilon+p)T^2}  \; K_3^\text{shear} - \frac{ \pi_\text{bulk}}{\zeta / \tau_\text{bulk}}\; K_1^\text{bulk}
\Big]
\nonumber\\
& -\frac{\partial \tau}{\partial\alpha}
\Big[
K^\text{eq}_2 + \frac{\pi^\eta_\eta}{2(\epsilon+p)T^2}  \; K_2^\text{shear}+ \frac{\pi^\phi_\phi}{2(\epsilon+p)T^2}  \; K_4^\text{shear} -  \frac{\pi_\text{bulk}}{\zeta/ \tau_\text{bulk}} \; K_2^\text{bulk}
\Big]
{\Bigg \}},
\label{eq:backgroundspectrumintermsofkernels}
\end{align}
Here $K_i^\text{eq}(p_{\rm T}, u^r), K_i^\text{shear}(p_{\rm T}, u^r)$ and
$K_i^\text{bulk}(p_{\rm T}, u^r)$ are rapidity and azimuthal angle integrated decay
kernels~\cite{Mazeliauskas:2018irt}. The kernels have implicit dependence on
scalars like freeze-out temperature or decay constants which do not vary on the
freeze-out surface.
The spectra of pions, kaons and protons as calculated with \Eq{eq:backgroundspectrumintermsofkernels} can then be compared to the experimentally measured $p_{\rm T}$ differential spectra of identified hadrons.

For the calculation of freeze-out kernels in \Eq{eq:backgroundspectrumintermsofkernels}, we use the publicly available code \textsc{FastReso} to perform strong
and electromagnetic decays of unstable hadrons\footnote{The feed-down  from weak decays of  $\Lambda$, $\Xi$ and $\Omega$ is not included in accordance with experimental procedure. We neglect resonance spectral widths and perform only the allowed 2-body and 3-body decays.} up to mass $m \approx
3\,\text{GeV}$. 
We use the list of $\sim 700$ resonances from refs.~\cite{Alba:2017mqu, Alba:2017hhe, parotto_private}, which is
based on all listed states (also less well established states) in the Particle Data Group 2016 publication~\cite{Patrignani:2016xqp}. This is approximately twice the number of resonances used previously~\cite{Mazeliauskas:2018irt}.
To perform a scan in freeze-out temperature, we varied it in the range $T_\text{fo}\in[130,180]\,\text{MeV}$ with $0.5\,\text{MeV}$ increments and zero baryon chemical potential. The transverse momentum $p_{\rm T}$ (in GeV) and the radial fluid velocity  $u^r$
have been discretized each on a 81 point non-linear grid in the range of $[0,3.5]$.

\section{Data analysis\label{sec:analysis}}

\subsection{Global fit procedure}

To summarize, our theoretical description has currently the  free parameters $\eta/s$, $(\zeta/s)_\text{max}$, the initialization time $\tau_0$, the freeze-out temperature $T_\text{fo}$ and they are assumed to be independent of the centrality class.
In addition, we have the 
normalization constants $\text{Norm}_i$ for the initial entropy profile which depend on the centrality classes.
In order to find the most likely model parameters, in this work we aim at fitting the $p_{\rm T}$-differential spectra of pions, kaons, and protons in five centrality intervals: 0--5\%, 5--10\%, 10--20\%, 20--30\% and 30--40\% for  Pb--Pb collisions at $\sqrt{s_{\rm NN}}$ = 2.76 TeV measured by the ALICE Collaboration~\cite{Abelev:2013vea}.
We choose to restrict to the soft particle momentum range  $p_{\rm T}<3\, \text{GeV}/c$, a region which is believed to be described by a fluid dynamic approximation to QCD dynamics. It is  sensitive to radial flow, the viscous transport coefficients and the initial conditions of the plasma~\cite{Gale:2013da, Bernhard:2016tnd, Teaney:2009qa, Dubla:2018czx}.

Nine model parameters are left free to vary simultaneously in specific intervals (see Table~\ref{tab:parranges}), in which the physical
values are expected to be located based on physical considerations and  previous work~\cite{Moreland:2014oya, Dubla:2018czx, Ryu:2017qzn}. Of course it is important to check {\it a posteriori} that the best fit values are indeed inside these intervals and not on its boundary (in the latter case one needs to allow for larger intervals).
 \begin{table}[ht!]
\begin{center}
    \begin{tabular}{ccccc}
    \hline
     \hline
     $\text{Norm}_i$  & $\tau_{0}$ (fm/$c$)& $\eta/s$ & $(\zeta/s)_\text{max}$ & $T_\text{fo}$ (MeV)\\ 
     50-67  & 0.1-0.6& 0.08-0.25 & 0.005-0.1 & 130-150\\ 
     \hline
      \hline
    \end{tabular}
  \caption{Ranges for  independently varied model parameters.}
    \label{tab:parranges}
\end{center}
\end{table} 

In order to determine which combination of the parameters provides the best description of the experimental data we search for the global minimum of

\begin{equation}
  \chi^2 = \sum^{N}_{i=1}\frac{(x_i - y_i)^2}{\sigma^2_i},\label{eq:chi2}
\end{equation}
where $x_i$ is the experimental value of the transverse momentum spectrum at some $p_{\mathrm{T}}$ interval for a particular hadron species and centrality class,  $y_i$ is the corresponding  model prediction (for a given set of model parameters) and  $\sigma_i=\sqrt{\sigma^2_{i,\text{sys}} + \sigma^2_{i,\text{stat}}}$ is the  sum (in quadrature) of the systematic and statistical uncertainties of the corresponding experimental data point. Let us remark here that we do not introduce a global theoretical uncertainty to all data points, as it was done in some previous studies \cite{Bernhard:2016tnd,Bernhard2019}.

The sum in \eqref{eq:chi2} goes over the five centrality classes (0--5\%, 5--10\%, 10--20\%, 20--30\% and 30--40\%), three particle species ($\pi, K,p$) and the number of $p_{\mathrm{T}}$ intervals in the fit range up to 3 GeV/$c$ ($N^\pi_{p_{\rm T}}=41$, $N^K_{p_{\rm T}}=36$ and $N^p_{p_{\rm T}}=34$). The total number of degrees of freedom is accordingly $N_\text{dof} = 555-9$. 
 
Note that the degree of correlation in the systematic uncertainties as a function of $p_{\rm T}$ in the experimental measurements is not reported and we do not consider such correlations in the fit. 
 
Furthermore, the simulations themselves might have considerable systematic uncertainties. For example, our model assumes a rather simple freeze-out picture without a detailed modeling of hadronic scatterings and the dissipative corrections to the single-particle distribution functions on the freeze-out surface are arguably somewhat uncertain. Also, our model neglects currently a possible temperature dependence of the shear viscosity to entropy ratio. Independently from this, also completely new physics might affect the experimental data in the low transverse momentum regime, for example pion condensation, see \cite{Begun:2015ifa, Begun:2016cva} and references therein. It is hard to predict {\it a priori} the change to model results due to such effects but for the interpretation of results it is important to keep in mind that theoretical uncertainties exist.

It remains to find the values of the model parameters which correspond to the fit of the experimental measurements with minimum ${\chi}^{2}$. We have explored here different strategies. What works best eventually is to discretize the model parameters on a hypercubic lattice and to use
numerical interpolation between the lattice points.
This allows to determine the $\chi^2$ landscape systematically and with the necessary precision.

We discretized the parameter ranges by 10 equidistant values for each parameter, which correspond to $10^5$ different model calculations for each centrality class. Let us note here that, thanks to streamlined fluid dynamic evolution and resonance decay procedures in our framework, one model simulation for  a particular set of parameters takes only a few tens of seconds per centrality on a single core and even in the exhaustive search with $10^5$ simulations, the entire fit can be performed with a rate of 1 day/centrality class using a $\sim$ 100 core machine.

Once all $10^5$ simulations have been computed we use an order-7 spline interpolation\footnote{In languages like Mathematica or Scientific Python such multi-dimensional, higher order B-spline interpolation schemes are readily implemented.} between them and apply a numerical minimization technique to find the lowest value of ${\chi}^{2}$ and its position.
For this minimization we used a Minuit algorithm \cite{James:1994vla} to find the global minimum.
The best fit found gives a $\chi^2/N_\text{dof}=1.37$. As a check of the numerical interpolation, we have also calculated $\chi^2$ directly for this specific configuration and obtained a compatible result.

The best fit parameters obtained in this way are reported in Table \ref{tab:bestfit1.par}. With the choice of the \textsc{TrENTo} parameter $p$ = 0 we observe that the values of $\rm Norm_i$ depend only mildly on centrality (as observed previously~\cite{Bernhard:2016tnd}).

\begin{figure*}
\centering
\includegraphics[width=1.\linewidth]{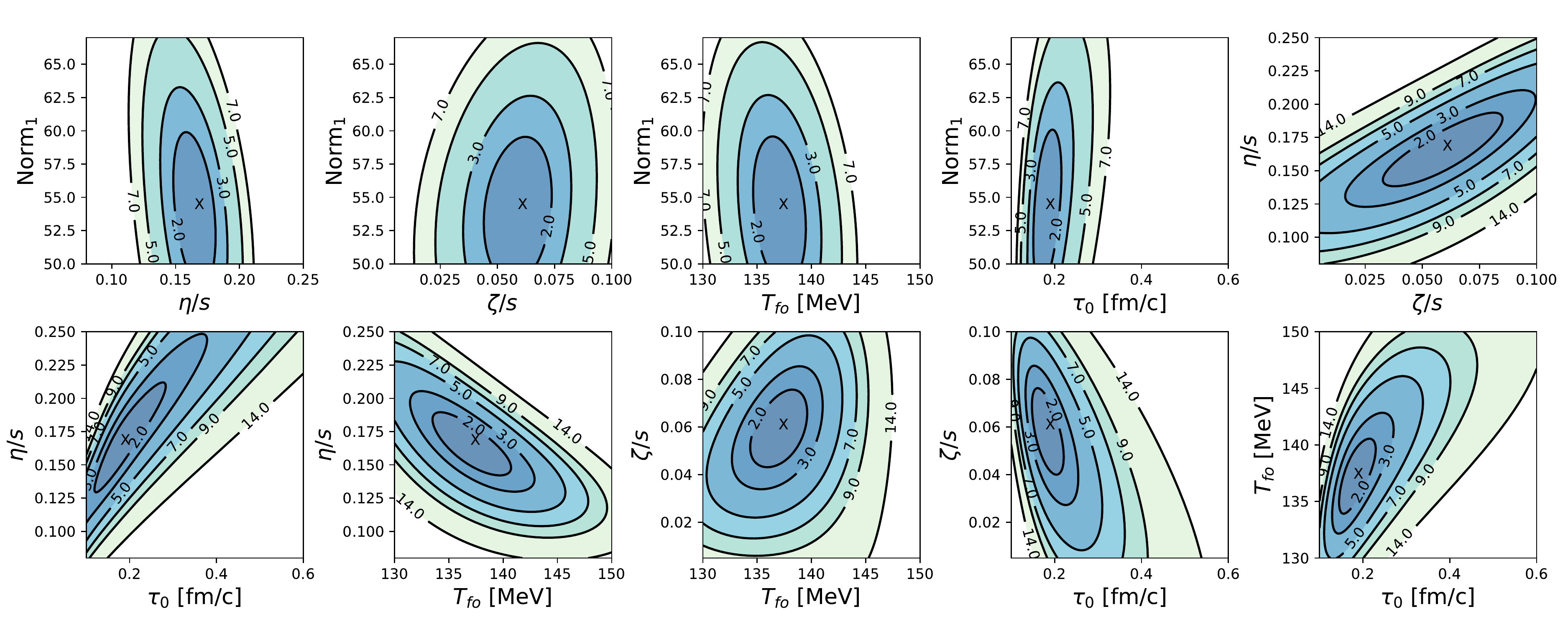}
\caption{Contour plots of ${\chi}^{2}/N_{\text{dof}}$ as a function of pairs of model parameters with all other parameters kept at the global minimum. The cross denotes the position of the minimum.}
\label{correlations}
\end{figure*}

\subsection{Uncertainties and correlations of model parameters}

In order to study correlations between pairs of model parameters according to their posterior probability distribution we use two methods. Firstly, two-dimensional slices of the nine dimensional ${\chi}^{2}$ landscape are computed with the remaining parameters kept at their global best fit (minimum $\chi^2$) value. The results are shown in  Fig. \ref{correlations}.

The first four panels on the top show the correlations of the initial entropy profile normalization $\text{Norm}_i$ with respect to the other four parameters. As an example we report the normalization for the centrality interval 0--5\%, $\text{Norm}_1$.
Thanks to the factored out scaling with initialization time $\tau_{0}$ in \Eq{eq:sinitial}, the different $\text{Norm}_i$ are observed to be almost independent from the other parameters. 
Only a rather weak correlation is observed between the initial normalization and $\eta$/s as well as $\tau_0$. This could be due to the combined effects of viscous entropy production at early times and the delayed generation of radial flow for larger $\tau_0$ values.
In the other six panels of Fig.~\ref{correlations} the correlations of the remaining parameter pairs are shown.
We see positive correlation between $(\zeta/s)_\text{max}$ and $\eta/s$, between $\tau_{0}$ and $\eta/s$ as well as between $\tau_{0}$ and $T_\text{fo}$. On the other hand, negative correlations are instead observed between $T_\text{fo}$  and $\eta/s$ as well as between $\tau_{0}$ and $(\zeta/s)_\text{max}$. Finally no strong and clear correlation is observed between $T_\text{fo}$ and $(\zeta/s)_\text{max}$. 

In order to quantify and supplement the information that is visually available in Fig.\ \ref{correlations}, we also determine numerically the form of $\chi^2$ as a function of the nine model parameters in the vicinity of the minimum. In terms of deviations from the best fit value $\Delta r$ = ($\Delta \text{Norm}_i$, $\Delta \tau_0/(\text{fm/c})$, $\Delta (\eta/s)$, $\Delta(\zeta/s)$, $\Delta T_\text{fo}/\text{MeV}$) we find
\begin{equation}
\chi^2 =\chi_{\text{min}}  + \sum_{i,j=1}^9 A_{ij} \Delta r_i \Delta r_j + \mathcal{O}(\Delta r^3). 
\label{eq:chi2quadratic}
\end{equation}
This information is interesting in particular because the probability for the correct fit parameters, given the experimental data we have analysed, is proportional to $e^{-\chi^2/2}$. The quadratic approximation to $\chi^2$ in eq.\ \eqref{eq:chi2quadratic} corresponds then to a Gaussian form of this so-called {\it posterior probability}. The diagonal values of the inverse matrix $A^{-1}$ can then formally be understood as variances of the fit parameters in this approximation to the posterior probability,
\begin{equation}
    \langle (\Delta r_j)^2 \rangle = (A^{-1})_{jj}.
    \label{eq:defDeltarj}
\end{equation}
Moreover, the matrix
\begin{equation}
    \rho_{ij} = \frac{(A^{-1})_{ij}}{\sqrt{(A^{-1})_{ii} (A^{-1})_{jj}}} = \frac{\langle \Delta r_i \Delta r_j \rangle}{\sqrt{\langle \Delta r_i^2 \rangle \langle \Delta r_j^2 \rangle}},
\label{eq:derhoij}
\end{equation}
quantifies correlations between the fitted parameters, again in a Gaussian approximation to the posterior distribution. Note that this information goes beyond what is visually available in Fig.\ \ref{correlations}. The latter shows two-dimensional sections through the $\chi^2$ landscape with the other parameters kept fixed. An expansion around the minimum gives the entries of the matrix $A_{ij}$. However, for the correlations as quantified in eq.\ \eqref{eq:derhoij} one needs actually the entries of the inverse matrix $A^{-1}$.

Note that these considerations assume that the uncertainties that enter Eq.\ \eqref{eq:chi2} are independent and normally distributed.

{
\def\arraystretch{0.5}
\begin{table*}
\centering
\begin{tabular}{ccccccccc} \hline \hline
$\text{Norm}_1$ & $\text{Norm}_2$ & $\text{Norm}_3$ & $\text{Norm}_4$ & $\text{Norm}_5$ & $\eta/s$ & $(\zeta/s)_\text{max}$ & $T_\text{fo}$ & $\tau_0$ \\
\hline
 1   &  0.89 &  0.89 &  0.89 &  0.89 & -0.72 & -0.78 &  0.67 &  0.49 \\
 0.89 &  1   &  0.89 &  0.89 &  0.89 & -0.72 & -0.77 &  0.66 &  0.49 \\
 0.89 &  0.89 &  1   &  0.89 &  0.89 & -0.72 & -0.77 &  0.65 &  0.5 \\
 0.89 &  0.89 &  0.89 &  1   &  0.89 & -0.71 & -0.76 &  0.64 &  0.5 \\
 0.89 &  0.89 &  0.89 &  0.89 &  1   & -0.71 & -0.76 &  0.63 &  0.49 \\ 
-0.72 & -0.72 & -0.72 & -0.71 & -0.71 &  1   &  0.97 & -0.88 &  0.13 \\
-0.78 & -0.77 & -0.77 & -0.76 & -0.76 &  0.97 &  1   & -0.85 & -0.01 \\
 0.67 &  0.66 &  0.65 &  0.64 &  0.63 & -0.88 & -0.85 &  1   &  0.01 \\
 0.49 &  0.49 &  0.5  &  0.5  &  0.49 &  0.13 & -0.01 &  0.01 &  1   \\
 \hline \hline
\end{tabular}
  \caption{Correlation matrix $\rho_{ij}$ between the fitted parameters in a Gaussian approximation to the posterior distribution as defined in eq.\ \eqref{eq:derhoij}.}
\label{rhoij}
\end{table*}
}

We show the resulting matrix $\rho_{ij}$ in Table\ \ref{rhoij}. The uncertainties on the model parameters according to eq.\ \eqref{eq:defDeltarj} are shown in Table \ref{tab:bestfit1.par} as uncertainties from the $\chi^2$ landscape. One remarks here that the latter are actually rather small. On the one side, this illustrates nicely that the experimental data are of high quality and have a high power to constrain theoretical models. On the other side, some remarks of caution about a too straight-forward interpretation are in order.

It is known that estimating model parameters including their uncertainty is difficult for situations with large $N_\text{dof}$ and when the minimum $\chi^2$ deviates substantially from its statistical expectation value (for a complete theoretical model) $\langle \chi^2 \rangle = N_\text{dof}$. This problem arises indeed for us when we attempt a global fit for the full range of transverse momenta and all five centrality classes with a single set of parameters. As a characteristic one may calculate the ``goodness of fit'' $Q=1-F_{\chi^2}(\chi^2, N_\text{dof})$ where $F_{\chi^2}(x,\nu)$ gives the cumulative $\chi^2$ distribution function with $\nu$ degrees of freedom. For the full global fit we find $ Q=1.8\times 10^{-8}$, which is indeed very small. This can be understood as the probability for the observed minimal $\chi^2$ given the data are correctly described by the model and all deviations from it arise indeed due to independent Gaussian fluctuations of the experimental data points. In other words, it is rather unlikely that the minimum $\chi^2=1.37\times N_\text{dof}$ we find (and in particular the deviation from the expectation value $\langle\chi^2 \rangle=N_\text{dof}$) arises due to statistical fluctuations only.

The fact that the goodness of fit is so small means that the theoretical model as it is currently implemented is in fact incomplete.
As we will see in the next section, the situation is not as bad, and our fluid model is certainly competitive with other attempts for theoretical descriptions, at least by visual inspection. Certain physics features might be missing, specifically in the low transverse momentum region for pions. Nevertheless, we should take the experimental data and the goodness of fit seriously. This leaves us with the problem to estimate the uncertainty of model parameters. 

\subsection{Estimation of systematic uncertainties}

In order to quantify how well the model parameters of the fluid description can actually be constrained from transverse momentum spectra, we cannot rely purely on the fit uncertainties, which are unrealistically small. As discussed above, the underlying reason is that the theoretical model is not complete. This can be seen directly from the goodness of fit estimate, but also indirectly from the fact that the outcome for the most likely model parameters depends on how the fits are being done in detail. In this subsection we will discuss this latter point, and estimate systematic uncertainties of the model parameters through variations of the fitting scheme.

\begin{figure}
\centering
\includegraphics[width=0.7\linewidth]{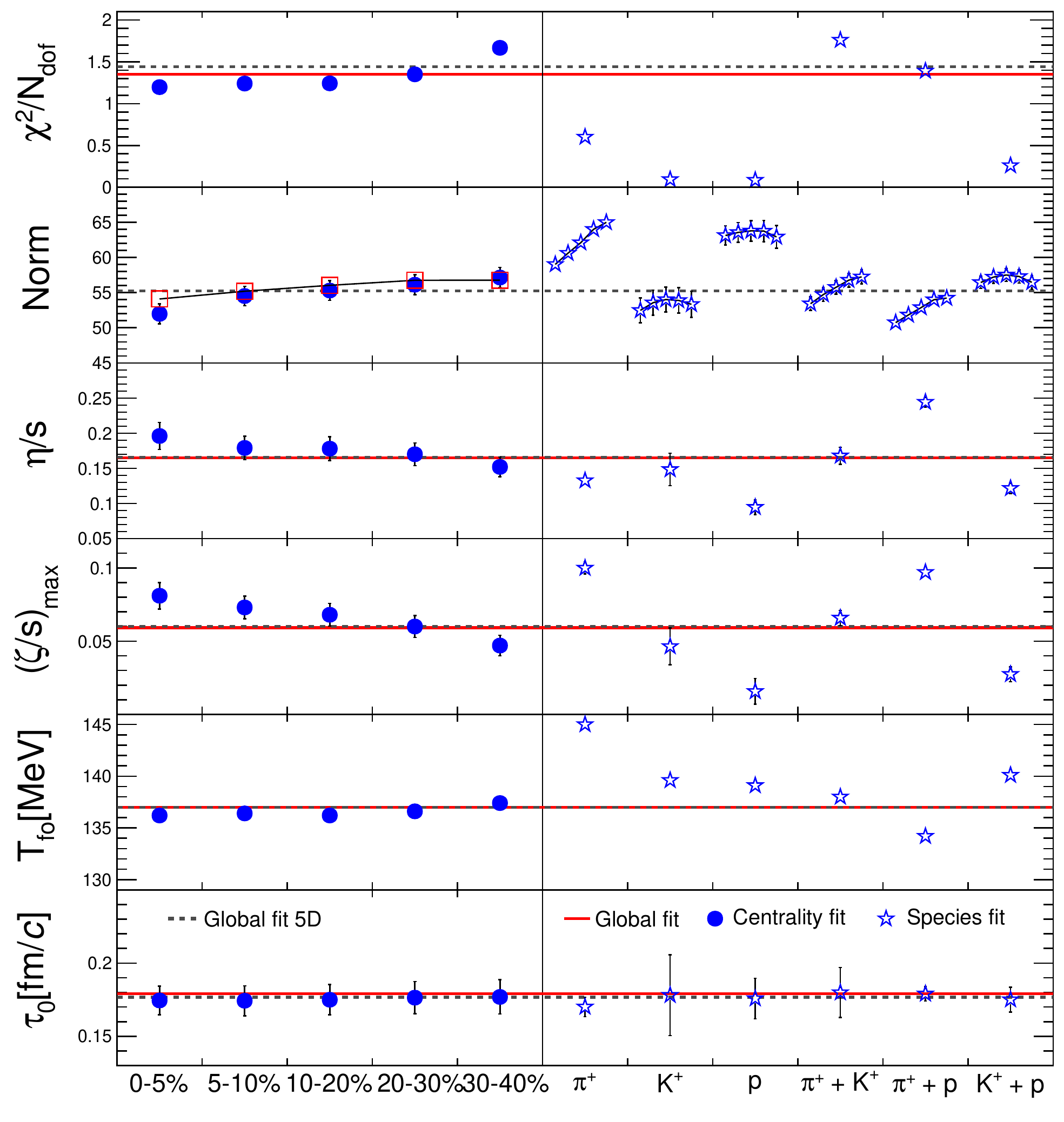}
\caption{Best fit model parameters and corresponding $\chi^2/N_\text{dof}$ obtained when different centrality classes are fitted separately (left side) and when the fit is restricted to kaons or kaons and pions (right side). We use these variations to estimate uncertainties of the best fit parameters as reported in Table~\ref{tab:bestfit1.par}.} 
\label{systematicc}
\end{figure}

The first check consists in fitting the five centrality classes separately and estimating the model parameters as a function of centrality. In addition to quantifying uncertainties, this test might also reflect possible temperature dependence of transport coefficients (specifically $\eta/s$). On the left hand side of Fig.\ \ref{systematicc} we show the result for the most likely model parameters when they are determined separately for the different centrality classes (full circles). The error bars illustrate the corresponding uncertainties according to eq.\ \eqref{eq:defDeltarj}, determined from the $\chi^2$ landscape. One finds that the variation arising from the centrality dependence is somewhat larger than the calculated fit uncertainties.

In a similar way, we also perform the fit separately restricted to single particle species, as well as restricted to two out of three particle species. This is done globally with respect to centrality. The results for the most likely model parameters obtained in this way are shown on the right hand side of Fig.\ \ref{systematicc} (open stars). One observes that the variations of fit parameters are here substantially larger than the statistical uncertainties estimated from the $\chi^2$ variation. On the other side, for the separate (single particle) fits of pions, kaons or protons we see that $\chi^2/N_\text{dof}$ drops below unity, which indicates the possibility of over-fitting.

Since the $\rm Norm_i$ does not show a significant centrality dependence, we have also tried to perform a global fit with 5 parameters, where only one common normalization for all centrality intervals is used. The results from this test is reported in with the black dashed line in Fig.\ \ref{systematicc} and as expected no significant variations are observed with respect to the default fit, however with a larger $\chi^2/N_\text{dof}$ = 1.47.

In Fig.~\ref{systematicc} the red 
lines represent the values obtained from the global fit reported in Table \ref{tab:bestfit1.par}. 
From the variations shown in Fig.~\ref{systematicc} we determine systematic uncertainties of the fitted model parameters and report them in Table \ref{tab:bestfit1.par} in the right-most column. Specifically, we take this uncertainty to be the maximal deviation seen in Fig.~\ref{systematicc} from the best fit parameter.

\begin{table}
\begin{center}
\begin{tabular}{c|c|c|c}
\hline
\hline
Model & Best fit & Uncertainty  & Uncertainty \\
parameter & value & from $\chi^2$ & from fit\\
& & landscape & variations \\
\hline
$\text{Norm}_1$ & 54.2 & $\pm$0.6 & -3.3, +9.0 \\
$\text{Norm}_2$ & 55.3 & $\pm$0.6 & -3.3, +8.4 \\
$\text{Norm}_3$ & 56.1 & $\pm$0.6 & -2.9, +7.7 \\
$\text{Norm}_4$ & 56.9 & $\pm$0.7 & -2.9, +7.2 \\
$\text{Norm}_5$ & 56.9 & $\pm$0.7 & -3.4, +6.2 \\
$\tau_{0}$ [fm/c] & 0.179 & $\pm$0.005 & -0.009, +0.001\\
$\eta/s$ & 0.164 & $\pm$0.007 & -0.07, +0.079 \\
$(\zeta/s)_\text{max}$ & 0.059 &  $\pm$0.003 & -0.043, 0.040\\
$T_\text{fo}$ [MeV] & 137.1 & $\pm$0.3 & -2.8, +8.0\\
\hline
\hline

\end{tabular}
\caption{Best fit parameters and their uncertainties determined  from the $\chi^2$ landscape through eq.\ \eqref{eq:defDeltarj}, and from the variation of the fitting procedure as reported in Fig.~\ref{systematicc}. For the global fit we find $\chi^2/N_\text{dof}=1.37$.}
\label{tab:bestfit1.par}
\end{center}
\end{table}

\section{Results and discussion\label{sec:results}}

The final step in the modeling workflow is to compute observables with best fit parameters, Table~\ref{tab:bestfit1.par}, and to make predictions for observables not used in the fit.
In this work, the simulations are performed for Pb--Pb collisions at $\sqrt{s_{\rm NN}}$ = 2.76 TeV for the centrality intervals 0--5\%, 5--10\%, 10--20\%, 20--30\% and 30--40\%.
First, we compare the fitted $p_{\rm T}$-differential spectra of identified hadrons to experimental measurements. Then we study the derived quantities, like the total multiplicities and mean-$p_{\rm T}$ for different hadron species.
Finally, we make model calculations for observables not used in the fit. Namely, we compute the $p_{\rm T}$ spectra for strange and multi-strange baryons at the same collision energy and centrality classes at $\sqrt{s_{\rm NN}}$ = 2.76 TeV and we make predictions for the pion, kaons, and protons spectra in Pb--Pb collisions at $\sqrt{s_{\rm NN}}$ = 5.02 TeV.

\subsection{Fitted  particle spectra of $\pi$, $K$, $p$\label{sec:spectra}}

\begin{figure*}
\centering
\includegraphics[width=\linewidth]{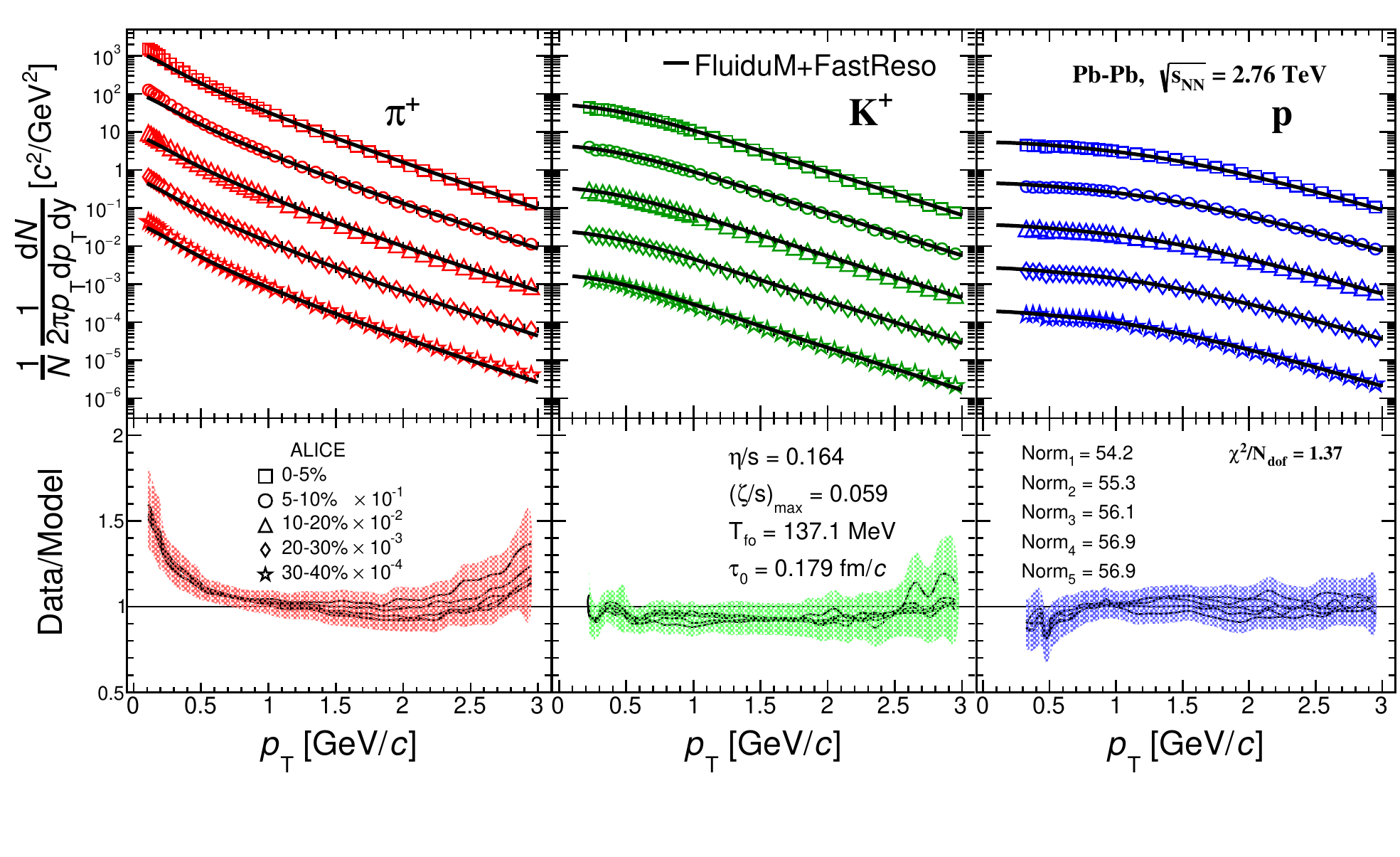}
\caption{Top: The best fit for $\pi, K, p$ spectra compared to the experimental data in five centrality classes in Pb--Pb collisions at   $\sqrt{s_{\rm NN}}$ = 2.76 TeV. Bottom: The data to model ratios. The shaded areas correspond to the sum in quadrature of the statistical and systematic experimental uncertainties. 
\label{bestfit}
}
\end{figure*}

In Fig.~\ref{bestfit} (top panels) we show the transverse momentum differential spectra of identified light hadrons $\pi$, $K$ and $p$ using our best fit  parameters ($\chi^2/N_\text{dof}=1.37$) listed in Table~\ref{tab:bestfit1.par} (lines) and we compare our results with the ALICE measurement (symbols). The bottom panels show the data to model ratio with shaded areas representing the combined experimental uncertainties.

Simulations are in overall good quantitative agreement with the experimental measurements. 
Kaon and proton spectra are reproduced within 10\%-20\% accuracy and within $3\sigma$ of experimental errors from the data or the entire $p_{\rm T} < 3\, (\text{GeV}/c)$ momentum range in all
the centrality classes. The pion spectra is reproduced well in a narrower $ 0.5< p_{\rm T} < 2.5\, (\text{GeV}/c)$ momentum range, while 
low-$p_{\mathrm{T}}$ pions are systematically underpredicted
and make major contributions
to the relatively large $\chi^2/N_\text{dof}=1.37$ in the fit. 
We checked that excluding soft pions from the fit results in a significantly smaller $\chi^2/N_\text{dof}\lesssim 0.6$ and the minimum moves out from the parameter ranges given in Table~\ref{tab:parranges}.
Such discrepancies in the pion spectra are well known
and have been observed both in hydrodynamic simulations~\cite{Song:2013qma,Alqahtani:2017tnq,Ryu:2017qzn, Dubla:2018czx} and blast-wave fits with resonance decays~\cite{Mazeliauskas:2019ifr,Melo:2019mpn}.

The enhancement of low-$p_{\mathrm T}$ pion spectra is typically attributed to the feed-down of resonance decays~\cite{Schnedermann:1993ws}.
However, even after we included a considerably larger set of primary resonances~\cite{Alba:2017mqu, Alba:2017hhe, parotto_private}, the agreement of soft pion spectra improved only marginally. 
Additional physics effects like finite widths of resonance decays~\cite{Huovinen:2016xxq},
the presence of pion condensation in heavy-ion collisions~\cite{Begun:2015ifa, Begun:2016cva} or going beyond linearised viscous corrections to the freeze-out spectra~\cite{Alqahtani:2017tnq} are being studied.

We would like to note here that our simulations show flat data to model ratio for protons within the uncertainties for the considered momentum range and in all centrality classes. However, a slight tendency towards over-predicting the low $p_{\mathrm T}$ protons is also observed. In this context it is interesting to note that in studies simulating a hadronic phase after chemical freeze-out~\cite{Ryu:2017qzn,Dubla:2018czx}, 
protons have been observed to receive an additional boost, resulting in a harder spectrum.

\begin{figure}
\centering
\includegraphics[width=0.49\linewidth]{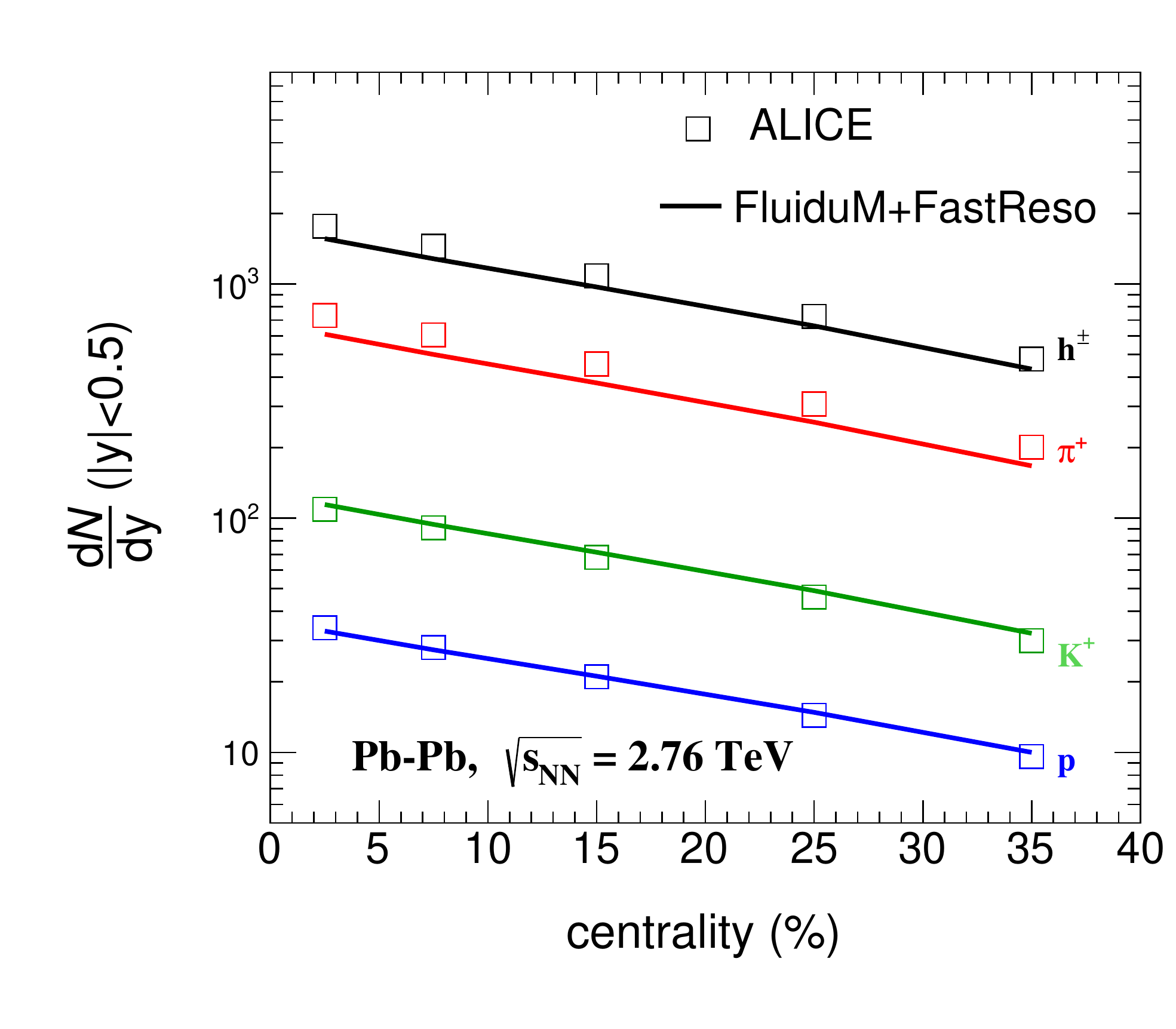} \hfill
\includegraphics[width=0.49\linewidth]{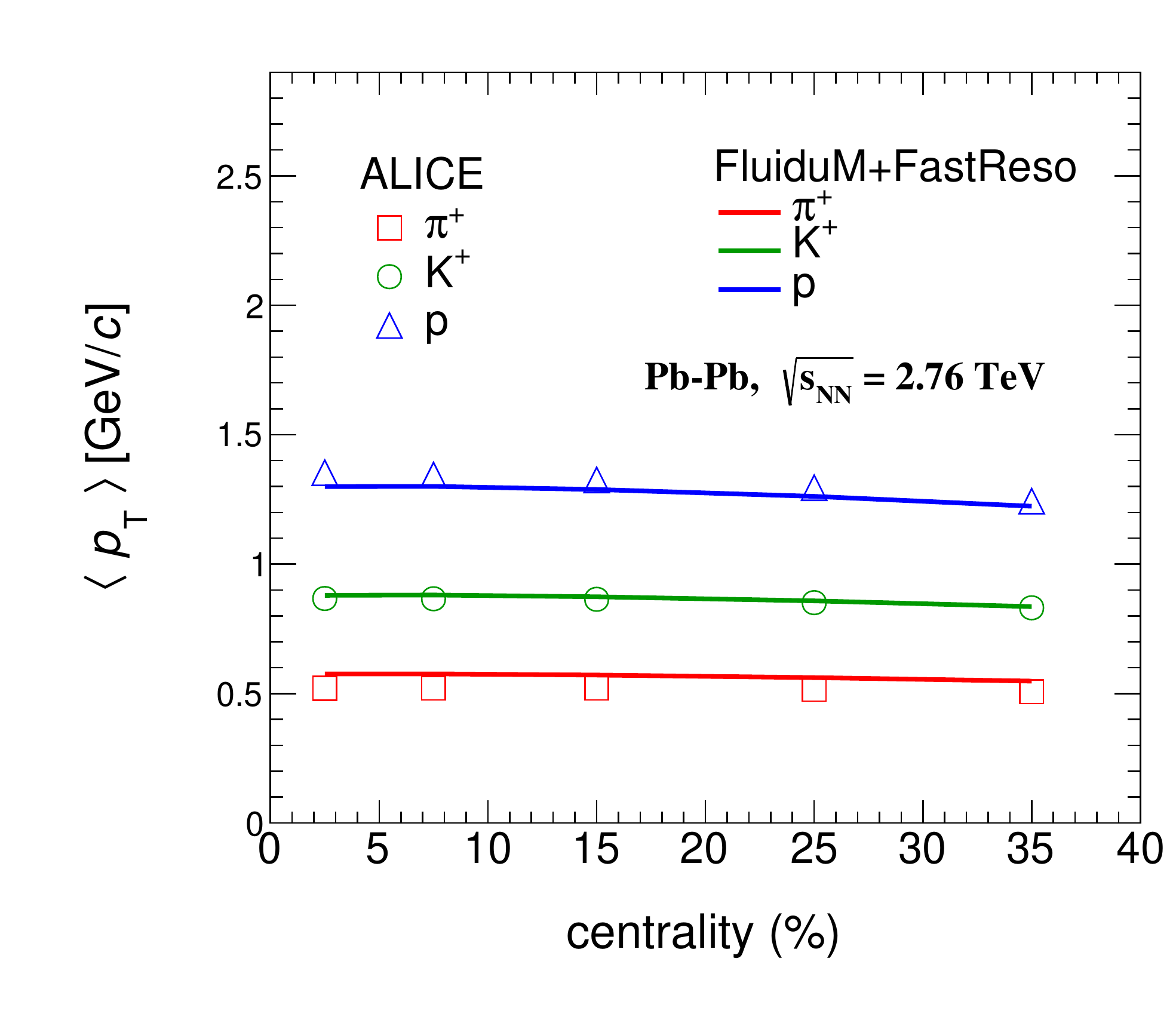} 
\caption{Mid-rapidity densities  d$N/$d$y$ (for $|y|<$ 0.5) for charged hadrons $h^\pm$, pions $\pi^+$, kaons $K^+$ and protons $p$ (top panel) and mean--$p_\text{T}$ (bottom panel) as functions of centrality from the calculation in comparison with the ALICE experimental measurements~\cite{Abelev:2013vea, alicemult}.}
\label{yield_and_meanpt}
\end{figure}

In addition to the data to model comparison of partice spectra, we can compute
other derived observables: particle multiplicity and mean $p_{\rm T}$.
In the top panel of Fig.~\ref{yield_and_meanpt} we compare our results of total charged and identified 
particle multiplicities at mid-rapidity as a function of collision centrality for pions, kaons, and protons with the ALICE measurements~\cite{Abelev:2013vea}. Our simulations give a reasonably good description of the centrality dependence of the charged hadron multiplicity. However, also in this case we see a tension with the pion and total charged hadron yields, especially in most central collisions, which is a clear reflection of the underestimation of the low $p_{\rm T}$ pion spectra observed in Fig.~\ref{bestfit}. 
In the bottom panel of Fig.~\ref{yield_and_meanpt} we compare the mean transverse momentum $\langle p_{\rm T} \rangle$, for pions, kaons, and protons as a function of centrality between our simulations and the experiment~\cite{Abelev:2013vea}. While $\langle p_{\rm T} \rangle$ of kaons agrees very well with the experimental measurements, 
the $\langle p_{\rm T} \rangle$ of pions and protons show some residual deviations. For the pions this is a reflection of the deviation between model and data in the transverse momentum spectrum below $p_{\rm T}$ = 0.5 GeV/$c$, which results in a slightly larger $\langle p_{\rm T} \rangle$ for pions in our model.
As for the protons, the slight discrepancy could be due to the absence of an hadronic phase between chemical and kinetic freeze-out in our model. 
We note that similar discrepancies are observed in other hydrodynamic simulations~\cite{Ryu:2017qzn, Alqahtani:2017tnq} and none appears able to reproduce data within the very small experimental uncertainties.

To our best knowledge no recent heavy-ion simulations (including our own presented here) are able to produce a uniformly good description of identified particle spectra from central to mid-central nucleus-nucleus collisions if experimental uncertainties are taken seriously. 
The pioneering studies of \cite{Bozek:2009dw} showed excellent agreement of identified particle spectra measured at RHIC with ideal hydrodynamic simulations, but
the agreement worsened when effects of viscosity were included.
In the \textsc{EKRT} model~\cite{Niemi:2015qia}, pion spectra are described well at the expense of over-predicted kaon and proton yields, which is in line with our finding when we attempt to fit only the pion spectra. In Ref.~\cite{Ryu:2017qzn} where the effect of both bulk viscosity and hadronic rescattering were studied, the data to model agreement is arguably on the same level as in our work, although we employ a single freeze-out approximation. We note here that the extensive Bayesian analyses of refs.~\cite{Bernhard:2016tnd,Bernhard2019} have concentrated on momentum integrated observables.
In summary, the excellent quality of experimental data of identified particle spectra indicates the need of including additional physics in hydrodynamic simulations of heavy-ion collisions.

\subsection{Strange, multi-strange and energy dependence of particle spectra\label{sec:predictions}}
Having found the optimal parameters of our model,
%
many other observables, not used in the fit, can be directly predicted. This is an important step
in validating the physics picture behind the model.
Therefore we use the fluid dynamic evolution with the best fit parameters to compute 
the $p_{\rm T}$ spectra of strange and multi-strange hadrons ($\Lambda$, $\Xi$, $\Omega$) and compare the results
with the ALICE measurements \cite{Abelev:2013xaa, ABELEV:2013zaa}.

\begin{figure*}[ht!]
\centering
\includegraphics[width=0.8\linewidth]{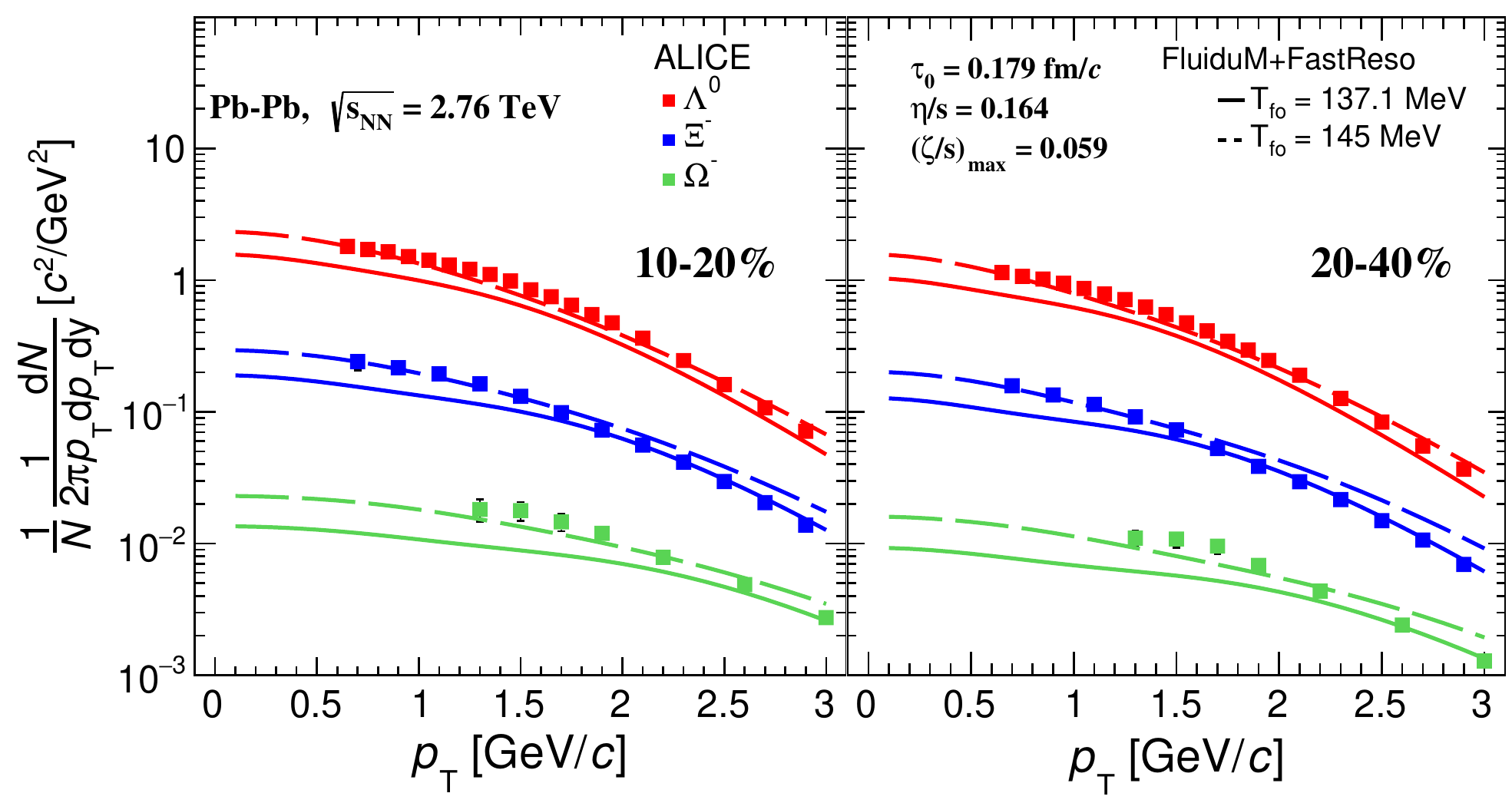}
\caption{ Differential $p_{\rm T}$ spectra
of strange and multi-strange baryons of Pb--Pb collisions with $\sqrt{s_\text{NN}}$ = 2.76 TeV. The normalization for both centrality intervals is taken from a global 5D fit with a value of 55.2.}
\label{LXO}
\end{figure*}

The comparison is shown in Fig.~\ref{LXO} for the 10--20\% (left panel) and 20--40\% (right panel) centrality intervals. 
From the comparison one can see that if the value of $T_\text{fo}= 137.1\,\text{MeV}$ is kept the same as obtained from the best fit (solid lines), the experimental $p_{\rm T}$-differential spectra of strange and multi-strange baryons are underestimated by the simulation. This effect is more pronounced for the $\Lambda$ baryons, which shows a $\sim 40\%-50\%$ discrepancy, while for the $\Xi$ and $\Omega$ the simulation and data tend to agree for $p_{\rm T}$ $>$ 2 GeV/$c$.
In previous work \cite{Ryu:2017qzn} it was observed that strange and multi-strange baryons are more sensitive to a change in the switching temperature from a fluid evolution to URQMD dynamics than pions, kaons and protons. In our case, if we increase the value of $T_\text{fo}$ to 145 MeV, while keeping all other parameters fixed, the simulation shows better agreement with data at low momentum, see in Fig.~\ref{LXO} (dashed lines), but  $\Xi$ is then over-predicted for $p_{\rm T}>2\,\text{GeV}/c$.

The tendency of strange and multi-strange hadrons preferring higher freeze-out temperatures~\cite{Bellwied:2018tkc,Bluhm:2018aei}, 
is sometimes used as an evidence for the scenario of sequential hadronization where the switching from quark to hadron degrees of
freedom occurs at different temperatures for different particle flavours~\cite{Bellwied:2013cta, Adamczyk:2017iwn,Alba:2020jir}. However, one should not discount the possibility that additional resonance feed-down might improve the agreement with data. Indeed, by approximately doubling the list of primary hadrons~\cite{Alba:2017mqu, Alba:2017hhe, parotto_private}, we observed a nearly 20\% increase in the feed-down for $\Lambda$ baryons compared to previous calculations~\cite{Mazeliauskas:2018irt}. Further extensions of decay channels and global fits including the strange particles would certainly reduce the apparent discrepancy.

\begin{figure}[ht!]
\centering
\includegraphics[width=0.49\linewidth]{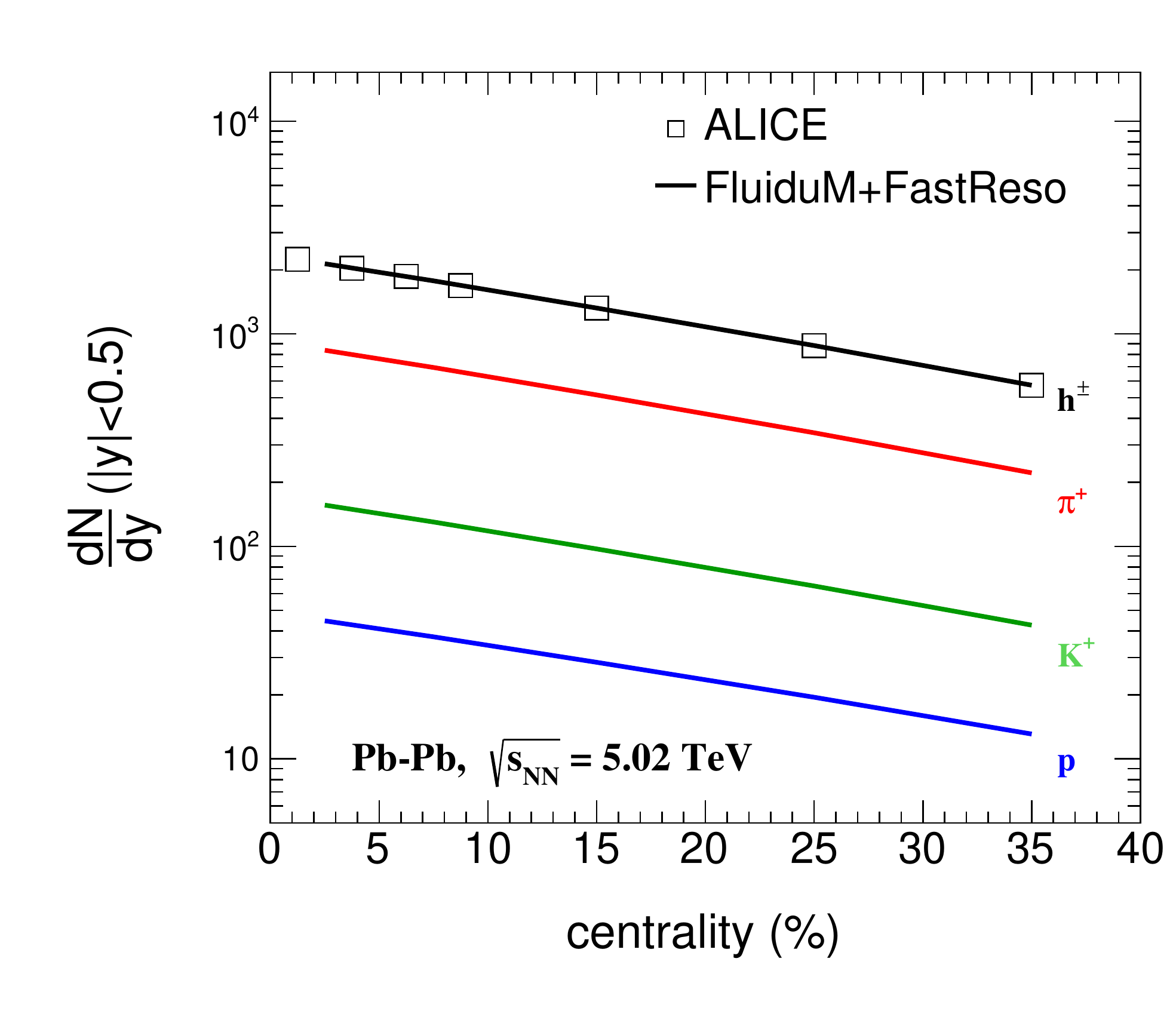}
\caption{Mid-rapidity densities dN/d$y$ ($\lvert y \rvert <$0.5) of charged hadrons as functions of centrality in Pb--Pb collisions at $\sqrt{s_\text{NN}}$ = 5.02 TeV in comparison with the ALICE measurements~\cite{Adam:2015ptt}. Prediction for the mid-rapidity densities dN/d$y$  ($\lvert y \rvert <$0.5)  of pions, kaon and protons are also reported.}
\label{5tevMult}
\end{figure}

\begin{figure*}[ht!]
\centering
\includegraphics[width=\linewidth]{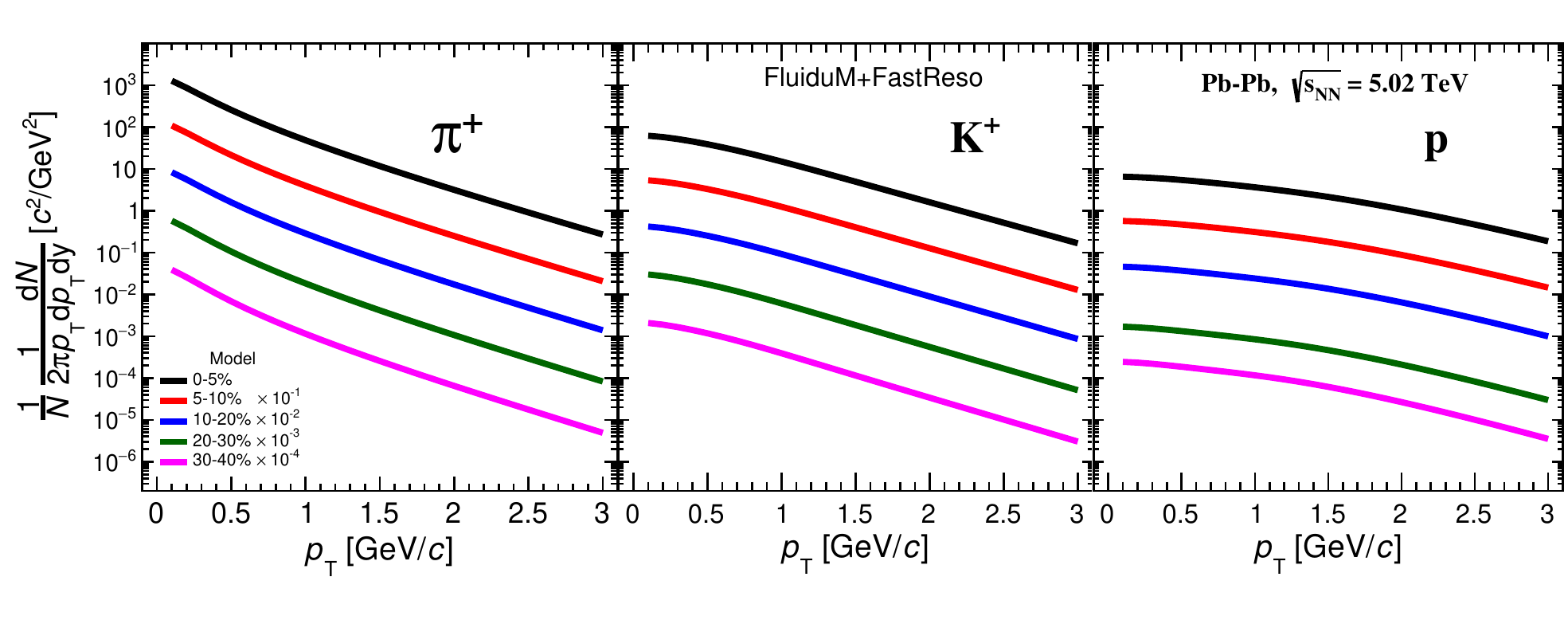}
\caption{Predictions for the $p_{\mathrm{T}}$--differential spectra  of pions (left panel), kaons (middle panel) and protons (right panel) in Pb--Pb collisions at $\sqrt{s_\text{NN}}$ = 5.02 TeV, in five centrality classes. }
\label{predictionspectra}
\end{figure*}

Finally, we can use our model to make predictions for the $p_{\mathrm{T}}$-differential spectra of pions, kaons and protons in Pb--Pb collisions at $\sqrt{s_\text{NN}}$ = 5.02 TeV. At higher collision energies, nuclei have more energy to deposit in the collision area, which ultimately results in an increased final particle multiplicity and higher initial QGP energy density. However as the increase of multiplicity is fractional, the  fundamental properties of the QGP are
not expected to change substantially and we can use the same best fit
model parameters to predict particle spectra at higher energies.
The only change made is the overall normalization $\text{Norm}_i$ of the initial entropy density profile.
The normalization at
$\sqrt{s_\text{NN}}$ = 5.02 TeV Pb--Pb collisions is fixed by doing a fit to the published unidentified charged hadron multiplicity as a function of the collision centrality~\cite{Adam:2015ptt}\footnote{We performed the fit in the same centrality classes as used for  $\sqrt{s_\text{NN}}$ = 2.76 TeV Pb--Pb  by combining the ALICE measurement at $\sqrt{s_\text{NN}}$ = 5.02 TeV energy into larger centrality bins. The new normalization factors are correspondingly $\text{Norm}_i= 75.6, 78.1, 77.8, 76.8, 76.4$.}.
We report the result in Fig.~\ref{5tevMult}, together with the model calculations for integrated yields of pions, kaons and protons as a function of centrality.
The corresponding plots for the $p_{\mathrm{T}}$-differential spectra of pions, kaons and protons in Pb--Pb collisions at $\sqrt{s_\text{NN}}$ = 5.02 TeV are reported in Fig.~\ref{predictionspectra} for the centrality intervals 0--5\%, 5--10\%, 10--20\%, 20--30\% and 30--40\%.
The $p_{\mathrm{T}}$-spectra at $\sqrt{s_\text{NN}}$ = 5.02 TeV are higher and flatter than the ones at $\sqrt{s_\text{NN}}$ = 2.76 TeV, which illustrates that stronger radial flow has been developed in the systems with larger final multiplicities at the higher collision energy.

\section{Summary and conclusion\label{sec:discussion}}
In summary, we have performed a global fit of transverse momentum particle spectra for identified pions, kaons and protons in five centrality classes based on a relativistic fluid approximation to QCD dynamics including a realistic thermodynamic equation of state as well as shear and bulk viscous dissipation, see ref.\ \cite{Floerchinger:2018pje} for further details. We have taken experimental data points measured at $\sqrt{s_\text{NN}}=2.76$ TeV by the ALICE collaboration in the transverse momentum range $p_\text{T} < 3$ GeV/$c$ as well as their experimental uncertainty into account, and searched for the most likely value of open parameters of the theoretical model on this basis.

One immediate result is the outcome for the most likely model parameters. They are summarized in Table \ref{tab:bestfit1.par}. In particular, the initialization time of the fluid description comes out relatively low, $\tau_0=0.179$ fm/c. For the shear viscosity to entropy ratio we find $\eta/s=0.164$, and for the peak value of the bulk viscosity to entropy ratio $(\zeta/s)_\text{max}=0.059$. The combined chemical and kinetic freeze-out temperature is determined to be $137.1$ MeV.

Our best fit value for the shear viscosity to entropy density ratio $\eta=0.164$ is rather close to the findings of ref.\ \cite{Niemi:2015qia} but deviates somewhat from the result of ref.\ \cite{Bernhard2019} which reports a minimum value $(\eta/s)_\text{min}=0.085$ at $T=154$ MeV and a positive slope towards larger temperatures. Ref.\ \cite{Dubla:2018czx} also pointed towards such a small minimal value $(\eta/s)_\text{min}=0.08$. On the other side, the analysis of ref.\ \cite{Kurkela:2019kip} pointed towards values in the range of our finding, specifically $\eta/s\approx 3/(4\pi)$.

Our best fit value for the freeze-out temperature is lower than found in other studies. In particular, the statistical hadronization model fits to integrated  light and multi-strange hadrons find $T_\text{ch}=156.5\,\text{MeV}$ in 0-10\% centrality bin~\cite{Andronic:2017pug}. However, excluding the multi-strange particles from the fit, as also done in our work, lowers the statistical hadronisation model fit down to $\sim 145\,\text{MeV}$ (see the recent publication \cite{Alba:2020jir}).
In addition, we note that the inclusion of (admittedly poorly understood) viscous corrections to the freeze-out distribution affects the best fit value of the freeze-out temperature. We checked that without these corrections, the best fit value indeed increases to $\sim 145\,\text{MeV}$. Finally, the number of included resonances also impact the optimal freeze-out temperature. Therefore such systematic differences in the modelling of hadronic freeze-out must be kept in mind when comparing different studies.

Moreover, from a quadratic expansion of $\chi^2$, corresponding to a Gaussian approximation to the posterior probability of the model parameters, we determine also their uncertainties as well as their correlation matrix, see Tables \ref{rhoij} and \ref{tab:bestfit1.par}. Note that in contrast to the Bayesian approach followed in refs.\ \cite{Bernhard:2016tnd,Bernhard2019}, our method to characterize the likelihood of the model parameters is independent of the parameter windows chosen as a prior. It is a local characterization using only the shape of the $\chi^2$ landscape in the vicinity of the minimum itself. Let us also emphasize that we take the entire form of the transverse momentum dependent particle spectra -- as well as the reported experimental uncertainties -- into account and not only integrated quantities such as total multiplicities or mean transverse momentum. 

From table \ref{tab:bestfit1.par} it becomes apparent that these uncertainties extracted from the $\chi^2$ landscape are rather small. This underlines the quality of the experimental data and their high power to constrain theoretical models. However, one must also say that the best fitting model parameters lead to $\chi^2/N_\text{dof}=1.37$ with $N_\text{dof}=546$. The deviation from the expectation value $\langle \chi^2 \rangle = N_\text{dof}$ is actually relatively large, which strictly speaking, implies that it is very unlikely that the current theoretical model correctly describes all of the observed physics. In other words, the residual deviations in Fig.\ \ref{bestfit} are statistically significant. The tension concerns in particular pions in the region of low transverse momenta. We may speculate which physics effect our model is missing. 

One possibility that comes to mind is that contributions from the feed down of decaying resonances have for some reason been underestimated. We have checked this possibility by doing our calculation with two different sets of hadronic resonances. While the current implementation uses the rather large set of $\sim 700$ resonances of ref.\ \cite{pdg}, we have also tried a smaller set based on an earlier listing ~\cite{Mazeliauskas:2018irt} and found the difference for the low-$p_\text{T}$ pions to be rather small. Of course, it cannot be fully excluded that an even larger set, or a more detailed description of the decay process~\cite{Lo:2017sux,Huovinen:2016xxq}, could remedy the problem.

Another interesting possibility is a non-thermal production mechanism for low-momentum pions such as from evolving coherent fields or condensates. An idea how this can happen in an out-of-equilibrium scenario is the one of a disoriented chiral condensate, see \cite{Mohanty:2005mv} for a review. Further work is needed to see whether such contributions from coherent fields and fluid dynamics can be reconciled.

Given that the current theoretical model is incomplete, it is rather difficult to determine its model parameters and the corresponding uncertainty. In particular, although straight-forward to calculate, the uncertainty from the $\chi^2$ landscape as quoted in Table \ref{tab:bestfit1.par} can {\it not} be taken as a complete estimate of uncertainty in a situation where the theoretical description is itself not yet complete. For this reason we have also studied how our best fit parameters change when the procedure for their determination is varied. Specifically, in Fig.\ \ref{systematicc} we show how the best fit parameters change if the fit is not done globally, i.e.\ for all centrality classes and all three particle species, but rather separately for individual centrality classes (and all three species), separately for pions, kaons and protons or for case in which only two species at a time are considered (but including all centrality intervals). One observes that this leads indeed to a sizeable variation of the model parameters and we estimate on this basis the uncertainties from fit variations in Table \ref{tab:bestfit1.par}. 

While in the present work we have focused on identified particle transverse momentum spectra, additional very interesting information is carried by harmonic flow coefficients and $n$-particle correlation functions. While they are sensitive to more detailed information from the initial state, their evolution is also highly sensitive to thermodynamic and transport properties \cite{Bernhard:2016tnd, Floerchinger:2013rya, Teaney:2003kp, Gale:2013da, Dubla:2018czx, Ryu:2017qzn, Song:2013qma, Niemi:2015qia, Bernhard2019}. Our theoretical framework \cite{Floerchinger:2018pje, Mazeliauskas:2018irt} has been developed also to describe those, and we plan to extend our theory-experiment comparison in this direction.

In conclusion we find that a fluid dynamic description of transverse momentum spectra for identified pions, kaons and protons works reasonably but with statistically significant residuals. The experimental data are now of a rather high quality and we expect that they will indeed allow to find a more complete theoretical description in the future. 

\section*{Acknowledgement}
The authors thank Paolo Parotto for sharing his PDG2016 resonance and decay lists.
This work is part of and supported by the DFG Collaborative Research Centre "SFB 1225 (ISOQUANT)". 
A.D. is partially supported by the Netherlands Organisation for Scientific Research (NWO) under
the grant 19DRDN011, VI.Veni.192.039.
Computational resources have been provided by the GSI Helmholtzzentrum f{\"u}r  Schwerionenforschung.

\bibliography{bib_hydro}
\bibliographystyle{JHEP}

\end{document}